\documentclass[12pt]{JHEP3}
\usepackage{bm}
\usepackage{amssymb,amsmath,amsthm}
\usepackage{mathrsfs} 
\def\TODAY{17 May 2009; 20 May 2009; 29 June 2009; 2 Sept 2009}
\title{\bf Quantum gravity without Lorentz invariance}
\author{Thomas P. Sotiriou\\
Department of Applied Mathematics and Theoretical Physics\\
Centre for Mathematical Sciences\\
University of Cambridge, Wilberforce Road, Cambridge CB3 0WA, UK\\
E-mail: \email{T.Sotiriou@damtp.cam.ac.uk}} 
\author{Matt Visser\\ School of Mathematics, Statistics, and Operations Research,\\
Victoria University of Wellington, PO Box 600, Wellington, New Zealand\\
E-mail: \email{matt.visser@msor.vuw.ac.nz}}
\author{Silke Weinfurtner\\
Physics and Astronomy Department, \\
University of British Columbia, \\
Vancouver, Canada\\
E-mail: \email{silke@physics.ubc.ca}}

\abstract{
There has been a significant surge of interest in Ho\v{r}ava's model for
3+1 dimensional quantum gravity, this model being based on anisotropic
scaling at a $z=3$  Lifshitz point. Ho\v{r}ava's model, and its variants,
 show dramatically improved ultra-violet behaviour at the cost of
exhibiting violation of Lorentz invariance at ultra-high
momenta. Following up on our earlier note, Phys.~Rev.~Lett.~{102} (2009) 251601 [arXiv:0904.4464
  [hep-th]], we discuss in more detail  our variant of Ho\v{r}ava's
model. In contrast to Ho\v{r}ava's original model, we abandon ``detailed
balance'' and restore parity invariance. We retain, however,  Ho\v{r}ava's
``projectability condition'' and explore its implications.  Under these conditions, we explicitly exhibit the most general model, and extract the full classical equations of motion in ADM form.  We analyze both spin-2 and spin-0 graviton propagators around flat Minkowski space. We furthermore analyze the classical evolution of FLRW cosmologies in this model, demonstrating that the higher-derivative spatial curvature terms can be used to mimic radiation fluid and stiff matter. We conclude with some observations concerning future prospects.

\TODAY; \LaTeX-ed \today.}
\begin{document}
\def\lint{\hbox{\Large $\displaystyle\int$}} 
\def\hint{\hbox{\Huge $\displaystyle\int$}}  
\def\d{{\mathrm{d}}}
\newcommand{\scri}{\mathscr{I}}
\newcommand{\sun}{\ensuremath{\odot}}
\def\d{{\mathrm{d}}}
\def\J{{\mathscr{J}}}
\def\L{{\mathscr{L}}}
\def\H{{\mathscr{H}}}
\def\T{{\mathscr{T}}}
\def\V{{\mathscr{V}}}
\section{Introduction}

Ho\v{r}ava's  ``Lifschitz point gravity''~\cite{Horava, Horava2, Horava3},  a field theoretic quantum gravity model based on ``anisotropic scaling'' of the space and time dimensions, has recently attracted a tremendous amount of attention.  Compared to traditional approaches, this model exhibits a vastly improved ultra-violet behavior --- it is certainly power-counting renormalizable, and arguably actually finite~\cite{Horava, Horava2, Horava3}. 
Early discussion regarding this model includes~\cite{LSB-regulator,
  Takahashi-Soda, cosmology1, cosmology2, Pope, Mukohyama,
  Brandenberger, Nikolic, Nastase, Cai, Cai2, Volovik, Piao, Gao,
  Colgain, 3+1, Chen, Mukohyama2, Myung, Cai3, Orlando, Gao2, Nishioka, 
  Kehagias, Rama, Cai4, Ghodsi, Mann, Chen2, Konoplya, Moffat, Chen3,
  Chen4}.  In particular, while Ho\v{r}ava's  specific model of
``Lifschitz point gravity'' as described in~\cite{Horava, Horava2,
  Horava3} has very many desirable formal features,  in its original
incarnation one appears to be forced to accept a non-zero cosmological
constant of the wrong sign to be compatible with
observation~\cite{Nastase, 3+1}. Additionally, one is forced to accept
intrinsic parity violation in the purely gravitational sector of 
  the  model \cite{Horava, Takahashi-Soda}. 

This naturally leads one to ask (and this is perhaps the key question from the quantum-field-theorist's point of view), just how important is ``detailed balance''? Is it an essential feature of the model, or is it just a simplifying assumption? Are there more general models that let you tune the Newton constant and cosmological constant independently? Can explicit parity violation in the pure gravity sector be eliminated?
Below, we shall provide more details concerning a variant of Ho\v{r}ava's model that is much better behaved in both these regards~\cite{3+1}, and which appears (at least superficially) to be phenomenologically viable.

Ho\v{r}ava has also introduced an explicit constraint which he calls the ``projectability condition''~\cite{Horava, Horava2, Horava3}. From a general relativists' point of view, this is the condition that a certain part of the space-time metric, the ``lapse function'', can be set globally to unity. Although at first glance this seems a significant constraint, the most important key solutions of the vacuum and cosmological Einstein equations (the Schwarzschild, Reissner--Nordstrom, Kerr, Kerr--Newman, Friedmann--Lemaitre--Robertson--Walker spacetimes) can all be put into this form --- at least for the physically interesting parts of those spacetimes~\cite{3+1, Living, Visser-on-Kerr}.  The key question (now from the point of view of a general relativist) is this: Is this ``projectability condition'' an essential feature of the model, or is it just a simplifying assumption?

A further issue is that Ho\v{r}ava's toy model  contains a spin-0 scalar graviton in addition to the standard spin-2 tensor graviton~\cite{Horava, Horava2, Horava3}. Phenomenologically, this is potentially risky, and might, for instance, run into constraints from the gravity-wave-dominated evolution of binary pulsar systems. Should that scalar mode be tuned to zero? Is there any symmetry that would protect this? (See for instance the discussion in~\cite{Cai3}, or more general comments within an ``emergent gravity'' framework~\cite{Gu,Xu}.)  Moreover, the toy model is purely gravitational, and it will need to be investigated carefully just how to embed matter (and eventually the standard model of particle physics) within it~\cite{cosmology1, cosmology2, Chen}. Because the gravitational field is still completely geometrical, albeit with a preferred frame, it might be that there is no intrinsic difficulty in maintaining a ÔuniversalÕ coupling to the gravitational field. Would there be non-zero signals in E\"o{}tv\"o{}s-type experiments? Apart from these questions, the community has already  begun to look at such things as the possible generation of chiral gravitational waves~\cite{Takahashi-Soda}, possible impacts on cosmological solutions and perturbations~\cite{cosmology1, cosmology2, Mukohyama, Brandenberger, Piao, Gao, Mukohyama2, Rama, Chen4}, possible modifications of black hole physics~\cite{Pope, Nastase, Cai, Cai2, Myung, Kehagias, Cai4, Mann, Chen2, Chen3, Larus},  the question of ``absolute time''~\cite{Nikolic}, ``emergent gravity''~\cite{LSB-regulator, Volovik, 3+1}, renormalizability~\cite{Horava, Horava2, Horava3, LSB-regulator, Orlando},  holography~\cite{Nishioka}, and more...

Following up on the brief sketch presented in~\cite{3+1}, we discuss in more detail some of the phenomenological implications of our extension of Ho\v{r}ava's model. In contrast to Ho\v{r}ava's original model, we abandon ``detailed balance'' and restore parity invariance. We retain, however,  Ho\v{r}ava's ``projectability condition''.  Under these conditions we explicitly exhibit all nine  spatial-curvature terms that contribute to the Lagrangian. As determined by power-counting, five of these operators are marginal (renormalizable) and four are relevant (super-renormalizable).  The classical limit of this extended model is phenomenologically much better behaved than Ho\v{r}ava's original model, as the Newton constant and cosmological constant can be independently adjusted to conform to observation.  Once the Planck scale and cosmological constant have been factored out, the model is described by eight independent dimensionless couplings, one of which is associated with the kinetic energy (and leads to a scalar graviton mode), while the other seven are related to the breaking of Lorentz invariance. We demonstrate how the Lorentz-breaking scales are related to the Planck scale, but need not be identical to the Planck scale, and sketch the first steps towards a detailed confrontation between this quantum gravity model and phenomenology. 

In particular we begin by extracting the full classical equations of
motion in ADM form, and after linearizing and gauge fixing, use them
to analyze both spin-2 tensor and spin-0 scalar graviton propagators
around flat Minkowski space, demonstrating that these two propagators
are sensitive to different subsets of the dimensionless coupling
constants. This analysis also yields the dispersion relations and
polarization states of classical gravitational radiation in the
weak-field regime.  Additionally, it provides useful insight in
  the relation between the ``projectability condition'' and the
  presence or absence of the spin-0 scalar graviton.  We furthermore analyze the classical evolution of FLRW cosmologies in this extended model, demonstrating that the modified Friedmann equations are sensitive to yet a third subset of the dimensionless coupling constants, and that the higher-derivative spatial curvature terms can be used to mimic both radiation fluid (``dark radiation'') and dark stiff matter. Thus different parts of the ``potential term'' govern distinct aspects of the phenomenology. We conclude with some observations concerning future prospects.

\section{The framework: Anisotropic scaling}
\subsection{Lapse, shift, and spatial metric}

To explain (our variant of) Ho\v{r}ava's approach~\cite{Horava, Horava2, Horava3}, the basic idea~\cite{LSB-regulator, 3+1} is to write the spacetime metric in ADM form
\begin{equation}
\d s^2 = - N^2 c^2 \d t^2 + g_{ij}(\d x^i - N^i \d t) (\d x^j - N^j \d t),
\end{equation}
and then, (adopting $\kappa$ as a placeholder symbol for some object with the dimensions of momentum), postulate that the engineering dimensions of  space and time are
\begin{equation}
[\d x] = [\kappa]^{-1}; \qquad [\d t] = [\kappa]^{-z}. 
\end{equation}
In condensed-matter language, this is typically referred to as ``anisotropic scaling''.
In particle-physics language, one is implicitly introducing a scale $Z$, with the physical dimensions $[Z]=[\d x]^z/[\d t]$, and using the theorists' prerogative to adopt units such that $Z\to1$.  Ultimately we shall interpret this scale $Z$ in terms of the Planck scale, and several closely related Lorentz-symmetry breaking scales. If one prefers to characterize this scale in terms of a momentum $\zeta$, then we have $Z=\zeta^{-z+1}\, c$, and we see that in order for dimensional analysis to be useful one cannot simultaneously set both $Z\to1$ and $c\to1$.  (Attempting to set both $Z\to1$ and $c\to1$ forces both $\d x$ and $\d t$ to be dimensionless, which then renders dimensional analysis utterly impotent, and destroys the ability to perform the desired ``power-counting'' analysis.) 
Consequently in these ``theoretician's units'' ($Z\to1$) one \emph{must} have
\begin{equation}
[N^i] = [c]  = {[\d x]\over[\d t]} = [\kappa]^{z-1},
\end{equation}
and one is free to additionally choose
\begin{equation}
[g_{ij}]  = [N] =  [1];\qquad  [\d s] = [\kappa]^{-1}.
\end{equation}
To minimize further algebraic manipulations, it is convenient to take the volume element to be
\begin{equation}
\d V_{d+1} =  \sqrt{g}\,  N \; \d^d x \; \d t; \qquad [\d V_{d+1}] = [\kappa]^{-d-z}.
\end{equation}
Note the \emph{absence} of any factor of $c$.  (This is simply a matter of convenience.) The resulting model will, by its very construction, violate Lorentz invariance;  the payoff however is greatly improved ultraviolet behaviour for the Feynman diagrams~\cite{Horava, Horava2, Horava3, LSB-regulator, 3+1, Orlando}, coupled with a well-behaved low-energy limit~\cite{Horava, Horava2, Horava3, LSB-regulator, 3+1}.

In fact we shall argue that a suitable extension~\cite{3+1} of the specific model presented by Ho\v{r}ava in~\cite{Horava} is (at least superficially) phenomenologically viable, and has a classical limit that is amenable to analysis in an ADM-like manner. Thus this is one of very few quantum gravity models that has any realistic hope of direct confrontation with experiment and observation. (When beginning the confrontation with phenomenology we will find it useful to go back to the more usual  ``physical units'' ($c\to1$) in which $Z\to \zeta^{-z+1}$.)

\subsection{Extrinsic and intrinsic curvatures}

Like the volume element, the extrinsic curvature is also most conveniently defined to not include any explicit factor of $c$:
\begin{equation}
K_{ij} = {1\over2N} \left\{ - \dot g_{ij} + \nabla_i N_j + \nabla_j N_i \right\}.
\end{equation}
Then $[N^i] =  {[\d x] / [\d t]} = [\kappa]^{z-1}$,
in agreement with the previous choices.  Furthermore
\begin{equation}
[K_{ij}] = {[g_{ij}]\over[N]  [\d t]} = [\kappa]^{z}.
\end{equation}
For the intrinsic curvature of the spatial slices we have
\begin{equation}
[g_{ij}] = [1]; \qquad [\Gamma^i{}_{jk}] = [\kappa];  \qquad [R^i{}_{jkl}] = [\kappa]^2,
\end{equation}
the key point being
\begin{equation}
[R^{ijkl}] = [\kappa]^{2};\quad [\nabla R^{ijkl}] = [\kappa]^{3};\quad [\nabla^2 R^{ijkl}] = [\kappa]^{4}.
\end{equation}

\section{Determining a suitable action functional}
\subsection{Kinetic term}
Consider the quantity
\begin{equation}
\T(K) = {g_K}     \left\{ (K^{ij} K_{ij} - K^2)+ \xi K^2 \right\}.
\end{equation}
(The general relativity kinetic term corresponds to the limit $\xi\to0$.)
Take the kinetic action to be
\begin{equation}
S_K =  \int \T(K) \; \sqrt{g} \; N\; \d^d x \; \d t.
\end{equation}
Again, note \emph{absence} of any factors of $c$. (This is again purely a matter of convenience to keep the dimensional analysis simple.) 
Then 
\begin{equation}
[S_K]  =   [g_K]  [\kappa]^{z-d}.
\end{equation}
Since the kinetic action is (by definition) chosen to be dimensionless, we have
\begin{equation}
[g_K] =  [\kappa]^{(d-z)}.
\end{equation} 
Note that the coupling constant $g_K$ is dimensionless exactly for 
\begin{equation}
d=z, 
\end{equation}
which is exactly the condition that was aimed for in~\cite{Horava}. In a simplified model based on scalar field self interactions, this is exactly the condition for well-behaved ultraviolet behaviour derived in~\cite{LSB-regulator}, and also the result obtained in~\cite{3+1}. Once we have set $d=z$ to make $g_K$ dimensionless, then provided $g_K$ is positive one can without loss of generality re-scale the time and/or space coordinates to set $g_K \to 1$. (A negative $g_K$ would ultimately lead to a wrong sign for the Newton constant, and $g_K=0$ is a physically diseased theory that has no kinetic energy terms.) Note that there is very little freedom in choosing the kinetic term: $\T(K)$ will be a generic feature of any Ho\v{r}ava-like model.

\subsection{Potential term}
Now consider a ``potential'' term 
\begin{equation}
S_\V =  -\int   \V(g) \; \d V_{d+1} =  - \int   \V(g) \; \sqrt{g} \; N \; \d^d x \; \d t,
\end{equation}
where $\V(g)$ is some scalar built out of the spatial metric and its spatial derivatives.
Again, note the \emph{absence} of any factors of $c$.
Then 
\begin{equation}
[S_\V] =  [\V(g)] \;  [\kappa]^{-d-z},
\end{equation}
whence
\begin{equation}
 [\V(g)] \to [\kappa]^{d+z}.
 \end{equation}
But to keep the kinetic coupling $g_K$ dimensionless we needed
$z\to d$. Therefore
\begin{equation}
 [\V(g)] \to [\kappa]^{2d}.
 \end{equation}
But $\V(g)$ must be built out of scalar invariants calculable in terms of the Riemann tensor and its derivatives, which tells us it must be constructible from objects of the form
\begin{equation}
\left\{  (\hbox{Riemann})^d, \; [(\nabla\hbox{Riemann})]^2 \hbox{(Riemann})^{d-3} , \;\hbox{etc...}   \right\}.
\end{equation}
In general, in $d+1$ dimensions this is a long but finite list. All of these theories should be well-behaved as quantum field theories~\cite{Horava, LSB-regulator, 3+1}.
In particular, since everything up to this point is valid for $d+1$ dimensions, these simple observations verify the main claims made in~\cite{Horava} regarding 3+1 dimensions, and shows that certain key aspects of that article nicely generalize to $d+1$ dimensions.

\subsection{Specializing to 3+1 dimensions}

In the specific case $d=3$ we have
\begin{equation}
 [\V(g)] \to [\kappa]^{6},
 \end{equation}
 and so obtain the much shorter specific list:
\begin{eqnarray}
&&
\Big\{  (\hbox{Riemann})^3,  [\nabla(\hbox{Riemann})]^2,    
(\hbox{Riemann})  \nabla^2(\hbox{Riemann}),   \nabla^4(\hbox{Riemann}) \Big\}.\qquad
\end{eqnarray}
But in 3 dimensions the Weyl tensor automatically vanishes, so we can always decompose the Riemann tensor into the Ricci tensor, Ricci scalar, plus the metric. Thus we need only consider the much simplified list:
\begin{equation}
\Big\{  (\hbox{Ricci})^3,  [\nabla(\hbox{Ricci})]^2,    
 (\hbox{Ricci})  \nabla^2(\hbox{Ricci}),   \nabla^4(\hbox{Ricci}) \Big\}. \quad
\end{equation}
We now consider a model containing all possible terms of this type, eliminating redundant terms using:
\begin{itemize}
\item Integration by parts and discarding surface terms.
\item Commutator identities.
\item Bianchi identities.
\item Special relations appropriate to 3 dimensions. \\
(Weyl vanishes; properties of  Cotton tensor.)
\end{itemize}
To keep the calculation tractable, (especially when it comes to
integration by parts), we impose Ho\v{r}ava's ``projectability'' condition
on the lapse function~\cite{Horava, 3+1}. This effectively is the
demand that the lapse $N(t)$ is a function of $t$ only, not a function
of position. (In particular, by re-parameterizing the time variable,
this means that without further loss of generality one can set
$N\to1$.) Besides simplifying the action, Ho\v{r}ava argues that 
  enforcing the projectability condition $N=N(t)$ might have other merits \cite{Horava2}: Since in his model the
  action is invariant only under foliation-preserving diffeomorphisms
  (see also section \ref{sec:gauge}), it would not be possible to use
  gauge transformations in order to set $N=1$ unless one had already enforced $N=N(t)$ at the outset. Taking also into account
  the special nature of time in the theory, this could cause
  technical difficulties in quantization~\cite{Horava, Horava2}.

It should be remarked that in standard general relativity this projectability condition can always be enforced \emph{locally} as a gauge choice. Furthermore for the most physically interesting solutions of general relativity it seems that this can always be done globally. For instance, for the Schwarzschild and Reissner--Nordstrom spacetimes this ``projectability'' condition holds globally (for the physically interesting region) in Painlev\'e--Gullstrand coordinates~\cite{3+1, Living}, while in the Kerr and Kerr-Newman spacetimes this condition holds globally (for the physically interesting $r>0$ region) in Doran coordinates~\cite{3+1, Visser-on-Kerr}. ``Projectability'' is also automatic for FLRW cosmologies. Thus in this sense the ``projectability condition'' does not seem to be a significant restriction on the physics.  

However, here we wish to enforce projectability at the level of the
action, and before any functional variation, and there is a  price to
pay for this: Firstly, not all gauges will be accessible to us at the
level of the field equations, as will be discussed shortly. Secondly,
as variation and gauge fixing do not necessarily commute the model we
are considering might not be the most general model with all possible
terms of dimension six. Nonetheless, one can expect that such a model
will not exhibit any qualitative phenomenological deviations from the
most general model, apart from those clearly related to the
projectability condition, which can be easily
distinguished. Therefore,  for the time being we shall take the
purely pragmatic approach of retaining projectability as a simplifying
assumption unless and until we are forced to abandon it. The use of
this ``projectability'' condition is a matter of some concern and
delicacy, as will be discussed later on.

After a brief calculation, we find that under these conditions there are only \emph{five} independent terms of dimension $[\kappa]^6$:
\begin{equation}
 R^3, \quad R \; R^{i}{}_{j} R^{j}{}_i, \quad  R^i{}_j R^j{}_k R^k{}_i; \quad 
R \; \nabla^2 R, \quad  \nabla_i R_{jk} \, \nabla^i R^{jk}.
\end{equation}
These terms are all marginal (renormalizable) by power counting~\cite{Horava,LSB-regulator,3+1}. In Ho\v{r}ava's article~\cite{Horava} only a particular linear combination of these five terms is considered: 
\begin{equation}
C^{i}{}_{j}  \; C^{j}{}_{i}.
\end{equation}
The restriction to this (Cotton)$^2$ term follows as a consequence of Ho\v{r}ava's, (to our minds), physically unnecessary ``detailed balance'' condition~\cite{Horava}.  Furthermore, as we shall soon see, this ``detailed balance'' restriction ultimately makes it difficult to set up a phenomenologically viable model based on Ho\v{r}ava's specific proposal~\cite{3+1}.

If we now additionally add all possible lower-dimension terms (relevant operators, super-renormalizable by power-counting) we obtain four additional operators:
\begin{equation}
[\kappa]^0: \; \;  1;
\qquad
[\kappa]^2: \; \;   R;
\qquad
[\kappa]^4: \; \;   R^2; \; \; R^{ij} R_{ij}.
\end{equation}
This now results in a potential $\V$ with nine terms and nine independent coupling constants.
In contrast, motivated by his ``detailed balance'' ansatz, Ho\v{r}ava~\cite{Horava} chooses a potential containing six terms (one of which is trivial) with only three independent coupling constants, of the form
\begin{equation}
\V_\mathrm{Horava}(g) = (\tilde g_2\,\hbox{Cotton}+\tilde g_1 \,\hbox{Einstein}+\tilde g_0\,\hbox{metric})^2.
\end{equation}
Note in particular that Ho\v{r}ava's approach includes the cross-term 
\begin{equation}
\hbox{(Cotton)}\times\hbox{(Einstein)} = \hbox{(Cotton)}\times\hbox{(Ricci)}.
\end{equation}
This is a $[\kappa]^5$ term which explicitly violates parity~\cite{Horava, Takahashi-Soda}, (and is the only parity-violating term in his model).  Because we have abandoned ``detailed balance'' we do not \emph{need} such a term, and we find it convenient to suppress it. It is worthwhile to emphasise that when trying to extend or modify Ho\v{r}ava's original model, it is the potential term that has most flexibility: the existence of \emph{some} $\V(g)$ is generic to Ho\v{r}ava-like models, but the details may vary from model to model. (For this reason many of the computations below are carried out for generic $\V(g)$.)

\subsection{Full classical action}
Assembling all the pieces we now have
\begin{equation}
S =  \int  \left[  \T(K) -  \V(g)  \right]  \sqrt{g}  \; N\; \d^3 x \; \d t,
\end{equation}
with
\begin{eqnarray}
\V(g)  &=&   g_0\, \zeta^6 +  g_1\, \zeta^4 \, R + g_2 \,\zeta^2\, R^2 + g_3 \, \zeta^2\,R_{ij} R^{ij} 
\nonumber\\
&&
+ g_4 \,R^3 + g_5 \,R (R_{ij} R^{ij}) + g_6\, R^i{}_j R^j{}_k R^k{}_i 
\nonumber\\
&&
+ g_7 \,R \nabla^2 R + g_8\,  \nabla_i R_{jk} \, \nabla^i R^{jk},
\end{eqnarray}
where we have introduced suitable factors of $\zeta$ to ensure the couplings $g_a$ are all dimensionless.
Assuming $g_1<0$, we can without loss of generality re-scale the time and space coordinates to set \emph{both} $g_K \to 1$ and $g_1\to -1$. 
The Einstein--Hilbert (+ cosmological constant) piece of the action is now (still in $Z\to1$ theoreticians' units) 
\begin{equation}
S_\mathrm{EH} =   \int   \left\{  (K^{ij} K_{ij} - K^2) +  \zeta^4 R  - g_0\, \zeta^6 \right\} \sqrt{g} \; N\; \d^3 x \; \d t,
\end{equation}
and the ``extra'' Lorentz-violating terms are controlled by a total of eight dimensionless coupling constants ($\xi$, $g_2$, \dots,  $g_8$)
\begin{eqnarray}
S_\mathrm{LV} &=&  \int   \Big\{ \xi\, K^2 -     g_2 \,\zeta^2\,R^2 -  g_3 \, \zeta^2\, R_{ij} R^{ij} 
\nonumber\\
&&
\qquad 
- g_4 \, R^3 - g_5 \, R (R_{ij} R^{ij}) - g_6 \, R^i{}_j R^j{}_k R^k{}_i 
\nonumber\\
&&
\qquad 
- g_7 \, R \nabla^2 R - g_8 \, \nabla_i R_{jk} \, \nabla^i R^{jk}
\Big\} \sqrt{g} \; N\; \d^d x \; \d t. \qquad
\end{eqnarray}
This is a perfectly reasonable classical Lorentz-violating theory of gravity, which we furthermore know has nice ultraviolet behaviour~\cite{Horava, LSB-regulator, 3+1}. Even classically, this model certainly deserves study in its own right. 

\subsection{Zeroth-order phenomenology: Recovering general relativity}

While the $Z\to1$ ``theoreticians'' units considered above have been most useful for power counting purposes, when it comes to phenomenological confrontation with observation it is much more useful to adapt more standard ``physical'' units ($c\to1$), in which $[\d x]=[\d t]$. The transformation to physical units is most easily accomplished by setting $(\d t)_{Z=1} \to \zeta^{-2} (\d t)_{c=1}$.  In these units the Einstein--Hilbert (+ cosmological constant) piece of the action becomes
\begin{equation}
S_\mathrm{EH} =   \zeta^2 \int   \left\{  (K^{ij} K_{ij} - K^2) +  R  -  g_0\, \zeta^2 \right\} \sqrt{g} \; N\; \d^3 x \; \d t.
\end{equation}
(See for instance equation (21.86) of MTW~\cite{MTW} for a comparison with standard general relativity.)
From this normalization of the Einstein--Hilbert term, we see that in these physical ($c\to1$) units
\begin{equation}
(16 \pi G_\mathrm{Newton})^{-1} = \zeta^2;  \quad \qquad \Lambda  = {g_0 \, \zeta^2\over2};
\end{equation}
so that $\zeta$ is identified as the Planck scale.  The cosmological constant is determined by the free parameter $g_0$, and observationally $g_0 \sim 10^{-123}$. 
In particular, the way we have set this up we are free to choose the Newton constant and cosmological constant independently (and so to be compatible with observation). In contrast, in the original model presented in~\cite{Horava}, a non-zero Newton constant \emph{requires} a non-zero cosmological constant, of the \emph{wrong sign} to be compatible with cosmological observations~\cite{Nastase, 3+1}.  (In a more realistic model including matter one would have to calculate the vacuum energy density and appropriately renormalize the cosmological constant, it would then be better to say that the nett renormalized value of  $g_0 \sim 10^{-123}$.)

\subsection{First-order phenomenology: Lorentz symmetry breaking}

In ``physical'' ($c\to1$) units, the ``extra'' Lorentz-violating terms become
\begin{eqnarray}
S_\mathrm{LV} &=&  \zeta^2 \int   \Big\{ \xi\, K^2 -     g_2 \,\zeta^{-2}\,R^2 -  g_3 \, \zeta^{-2}\, R_{ij} R^{ij} 
\nonumber\\
&&
\qquad 
- g_4 \,  \zeta^{-4}\,R^3 - g_5 \,\zeta^{-4}\, R (R_{ij} R^{ij})
\nonumber\\
&&
\qquad 
 - g_6 \,\zeta^{-4}\, R^i{}_j R^j{}_k R^k{}_i 
- g_7\,\zeta^{-4}\, R \nabla^2 R
\nonumber\\
&&
\qquad 
 - g_8 \,\zeta^{-4}\, \nabla_i R_{jk} \, \nabla^i R^{jk}
\Big\} \; \sqrt{g} \; N\; \d^d x \; \d t. \qquad
\end{eqnarray}
The extra Lorentz violating terms consist of one kinetic term, and seven higher-spatial-curvature terms.
The Lorentz violating term in the kinetic energy leads to an extra scalar mode for the graviton~\cite{Horava}, with fractional $O(\xi)$ effects at all momenta.  (See further discussion below.) Phenomenologically, this behaviour is potentially dangerous and should be carefully investigated. 
In contrast the  various Lorentz-violating terms in the potential become comparable to the spatial curvature term in the Einstein--Hilbert action only for physical momenta of order
\begin{equation}
\zeta_{\{2,3\}} = {\zeta\over\sqrt{ |g_{\{2,3\}}| }},  \qquad \zeta_{\{4,5,6,7,8\}} = {\zeta\over\sqrt[4]{ |g_{\{4,5,6,7,8\}}| }},
\end{equation}
or higher. 
Thus the higher-curvature terms are automatically suppressed as we go to low curvature (low momentum).  

Note that in this analysis we have now divorced the Planck scale $\zeta$ from the various Lorentz-breaking scales $\zeta_{\{2,3,4,5,6,7,8\}}$, and that we can drive the Lorentz breaking scale arbitrarily high by suitable adjustment of the dimensionless couplings $ g_{\{2,3\}}$ and  $g_{\{4,5,6,7,8\}}$. It is these pleasant properties that make the model phenomenologically viable --- at least at a superficial level --- and that encourage us to consider more detailed confrontation with experiment and observation. Since the ultraviolet dominant part of the Lorentz breaking is sixth order in momenta, it neatly evades all current bounds  on Lorentz symmetry breaking~\cite{Mattingly, Liberati, Planck}. At most one might hope to get some observational constraints on $ g_2$ and/or $ g_3$, (which lead to deviations from Lorentz invariance at fourth order in spatial momenta), but even those bounds would be rather weak.

\section{Classical equations of motion}
\subsection{Hamiltonian constraint}
Varying with respect to the lapse $N(t)$ one obtains the Hamiltionian constraint
\begin{equation}
H = \int \sqrt{g} \,\H(K,g) \,\d^3 x =\int \sqrt{g} \,\{ \T(K) + \V(g) \} \, \d^3 x = 0.
\end{equation}
The difference here compared to standard general relativity lies in:
\begin{enumerate}
\item  The $\xi$ term in the kinetic energy.
\item  The more complicated form of the potential $\V(g)$.
\item  Finally, because of the assumed ``projectability'' condition on the lapse $N(t)$ one cannot derive a super-Hamiltonian constraint, and must remain satisfied with this \emph{spatially integrated} Hamiltonian constraint.
\end{enumerate}
We emphasise that this Hamiltonian constraint will be \emph{generic} to all Ho\v{r}ava-like models as it really only depends on the ``projectability'' condition and is completely independent of the precise form of the potential $\V(g)$.  Furthermore, if (and only if) one relaxes the ``projectability'' condition and permits the lapse $N(t,x)$ to be an arbitrary function of space+time, would one then obtain a super-Hamiltonian constraint
\begin{equation}
\H(K,g) \equiv \T(K) + \V(g)  = 0.
\end{equation}

\subsection{Super-momentum constraint}

Varying with respect to the shift $N^i$ one obtains the super-momentum constraint 
\begin{equation}
\nabla_i \pi^{ij}= 0,
\end{equation}
where the super-momentum is
\begin{equation}
\pi^{ij} = {\partial [ N \T(K)] \over\partial \dot g_{ij} } =   - \left\{ K^{ij} - K g^{ij}  + \xi K g^{ij} \right\}.
\end{equation}
The difference here compared to standard general relativity is utterly minimal and lies solely in the $\xi$ term. (See for instance equation (21.91) of MTW~\cite{MTW} for a comparison with standard general relativity.)
We emphasise that this super-momentum constraint will be \emph{generic} to all Ho\v{r}ava-like models as it only depends on the form of the kinetic term $\T(K)$ and is completely independent of the precise form of the potential $\V(g)$. 

\subsection{Dynamical equation}
By varying with respect to $g_{ij}$ one now obtains
\begin{eqnarray}
\label{piequation}
{1 \over \sqrt{g}} \; \partial_t ( \sqrt{g} \; \pi^{ij} ) &=&  -2N \left\{ (K^2)^{ij} - K K^{ij} + \xi K K^{ij} \right\} 
\nonumber\\
&& 
+ {N\over2} \T(K)\; g^{ij}  + (\nabla_m N^m) \; \pi^{ij} + [\L_{\vec N} \pi]^{ij}
\nonumber\\
&& 
+{N\over \sqrt{g}}   {\delta S_\V \over\delta g_{ij} }.
\end{eqnarray}
This is very similar to standard general relativity (see for instance equation (21.115) of MTW~\cite{MTW} for a comparison): 
\begin{itemize}
\item There is a straightforward extra contribution coming from the $\xi$ term in the kinetic energy.
\item The only real subtlety lies in evaluating the ${\delta S_\V /\delta g_{ij} }$ terms.
\item We emphasise that most of the features of this dynamical equation will be \emph{generic} to all Ho\v{r}ava-like models.  The specific model dependence is confined to the ${\delta S_\V /\delta g_{ij} }$ term.
\item In contrast to standard general relativity there are no terms depending on the spatial gradients of the lapse function --- this is a side effect of the ``projectability'' condition. 
\end{itemize}
As usual $(K^2)^{ij}= K^{im} \, g_{mn} \, K^{nj} = K^{ik}\;K_k{}^j$.
Evaluating the  ${\delta S_\V /\delta g_{ij} }$ term for our specific variant of Ho\v{r}ava's model is somewhat tedious, but since in our model we know that $S_\V$ is the most general action one can build out of the metric using 0, 2, 4, or 6 derivatives we can (without calculation) 
deduce that the ``forcing term''
\begin{equation}
F^{ij} = {1\over \sqrt{g}}  \;  {\delta S_\V \over\delta g_{ij} }.
\end{equation}
is the most general symmetric conserved tensor one can build out of the metric and 0, 2, 4, or 6 derivatives.  Writing 
\begin{equation}
F^{ij} = \sum_{s=0}^8 g_s \;  \zeta^{n_s} \; (F_s)^{ij},
\end{equation}
where $n_s$ is an appropriate integer to get the dimensions correct,  
an explicit calculation identifies the following nine individual terms contributing to the overall forcing term: 
\begin{description}
\item[$g_0$:]     $ 1 \to $
\begin{equation}
  (F_0)_{ij} = -{1\over2} \; g_{ij}.
\end{equation}

\item[$g_1$:]     $ R \to  $
\begin{equation}
(F_1)_{ij} = G_{ij} .
\end{equation}

\item[$g_2$:]     $ R^2 \to  $
\begin{equation}
(F_2)_{ij} = 2R R_{ij} - {1\over2} R^2 g_{ij} - 2[\nabla_i \nabla_j - g_{ij} \nabla^2 ] R.
\end{equation}

\item[$g_3$:]    $ R_{mn} R^{mn} \to $ 
\begin{equation}
(F_3)_{ij} =   {3\over2} (R_{mn} R^{mn}) g_{ij} + 
\nabla^2 R_{ij}  + {1\over2} g_{ij} \nabla^2 R - \nabla_i \nabla_j R +   3 R R_{ij} - 4(R^2)_{ij} -  R^2 g_{ij}.
\end{equation}

\item[$g_4$:] $ R^3 \to $
\begin{equation}
(F_4)_{ij} = 3 R^2 R_{ij} - {1\over2} R^3 g_{ij} - 3 [\nabla_i \nabla_j - g_{ij} \nabla^2] R^2.
\end{equation}

\item[$g_5$:] $ R \; (R_{mn} R^{mn})  \to $
\begin{eqnarray}
 (F_5)_{ij} &=&    R_{ij}  (R_{mn} R^{mn})  + 2 R (R^2)_{ij} -{1\over2}  R\, (R_{mn} R^{mn} ) g_{ij} 
\\
&&
 + \left[ \nabla^2( R R_{ij})   +  \nabla_m \nabla_n ( R R^{mn} )   g_{ij} 
 - \nabla^k \nabla_i ( R R_{jk}) 
 -  \nabla^k \nabla_j  (R R_{ki}) \right]
\nonumber\\
&&
+ \left[ g_{ij} \nabla^2 + \nabla_i \nabla_j\right]  (R_{mn} R^{mn}) .
\nonumber
 \end{eqnarray}
 
\item[$g_6$:] $ R^m{}_n R^n{}_p R^p{}_m \to $
\begin{eqnarray}
 (F_6)_{ij}  &=& 3 (R^3)_{ij} - {1\over2} ( R^k{}_m R^m{}_n R^n{}_k ) g_{ij} 
   \\
&&  +{3\over2} \left[  \nabla^2 (R^2)_{ij} + \nabla_n \nabla_m (R^2)^{nm} g_{ij} 
 - 
 \nabla^n \nabla_i (R^2)_{jn} 
 -
 \nabla_n \nabla_j (R^2)_{in} \right].
 \nonumber
\end{eqnarray}

\item[$g_7$:] $ (\nabla R)^2 = (\nabla_m R) g^{mn} (\nabla _n R) \to $
 \begin{equation}
  (F_7)_{ij}  = - 2 [g_{ij} \nabla^2 - \nabla_i \nabla_j] \nabla^2 R  - 2 (\nabla^2 R) R_{ij} +  (\nabla_i R) \,(\nabla_j R) -  {1\over2}  (\nabla R)^2 g_{ij}.
 \end{equation}
 
 \item[$g_8$:] $ (\nabla_p R_{mn}) (\nabla^p R^{mn}) \to $
\begin{eqnarray}
 (F_8)_{ij} &=& 
  \nabla^4 R_{ij} + (\nabla_p \nabla_q \nabla^2 R^{pq})  g_{ij}  
  -  \nabla_p \nabla_i \nabla^2 R^{p}{}_{j} -   \nabla_p \nabla_j \nabla^2 R^{p}{}_{i} \qquad
 \\
 &&
 - (\nabla_i R^m{}_n)\,(  \nabla_j R^n{}_m)  - {1\over 2}   (\nabla_p R_{mn}) (\nabla^p R^{mn})  g_{ij} -  2 (\nabla^q R_i{}^{p}) (\nabla_q R_{jp}).
 \nonumber
 \end{eqnarray}
\end{description}
As usual we adopt the notation $(R^2)^{ij}=  R^{ik}\;R_k{}^j$, and  similarly $(R^3)^{ij}=  R^{ik}\; R_k{}^l \; R_l{}^k$. Furthermore $(\nabla R)^2 = (\nabla_i R) \, (\nabla^i R)$.
The first two terms above, $(F_0)_{ij}$ and $(F_1)_{ij}$, are exactly what one would expect for the 3+1 decomposition of standard Einstein gravity with a cosmological constant. The remaining seven forcing terms characterize violations of Lorentz invariance and from a QFT perspective are utterly essential for regulating the high-energy behaviour of the Feynman diagrams~\cite{Horava, LSB-regulator, 3+1}. From a classical perspective, these are just higher-(spatial)-curvature terms which are suppressed at low curvature by suitable powers of the relevant Lorentz-breaking scales $\zeta_{\{2,3,4,5,6,7,8\}}$.
The relevance of these observations is that the classical limit has now been cast into an ADM-like form, suitable, for instance, for detailed numerical investigations (and other purposes).

\section{Scalar and tensor graviton propagators around Minkowski space}
\def\zero{ {}^0 }
\def\one{  {}^1 }
\def\two{ {}^2 }
\subsection{Linearizing the equations of motion}
 
To extract the graviton propagator we will linearize the equations of
motion derived above. In the classical limit, this calculation will also yield the dispersion relations and polarization states for classical weak-field gravitational radiation.   For simplicity we will set the cosmological constant $g_0=0$ for this calculation and use flat spacetime as a background solution. We then have
\begin{equation}
\zero g_{ij} = \delta_{ij} ; \qquad \zero N_i = 0;  \qquad \zero N = 1.
\end{equation}
Now consider linearized perturbations
\begin{equation}
g_{ij} = \delta_{ij} + \epsilon \;h_{ij}; \qquad N_i =  \epsilon \; n_i; \qquad N = 1 + \epsilon \; n(t),
\end{equation}
then
\begin{equation}
g^{ij} = \delta^{ij} - \epsilon\; h^{ij} + O(\epsilon^2) ; \qquad N^i = \epsilon\; n^i+ O(\epsilon^2); \qquad N = 1 + \epsilon\; n(t).
\end{equation}
Now expand the extrinsic curvature $K_{ij}$, the conjugate momentum $\pi_{ij}$, the intrinsic Ricci curvature $R_{ij}$ and forcing term $F_{ij}$ as a series in $\epsilon$ of the form
\begin{equation}
X = \zero X + \epsilon \; \one X +O(\epsilon^2)
\end{equation}
In fact $K_{ij}$, $\pi_{ij}$, $R_{ij}$, and $F_{ij}$ all vanish at zeroth order and the first non-trivial contribution arises at order $\epsilon$.

\subsubsection{Hamiltonian constraint}

Since $K_{ij} = O(\epsilon)$, it follows that $\T(K)=O(\epsilon^2)$. Furthermore, explicit inspection of the potential shows that there is only one non-trivial term
\begin{equation}
\V(g) =   \epsilon \; g_1 \; (\one R) + O(\epsilon^2).
\end{equation}
So the linearized Hamiltonian constraint is
\begin{equation}
\label{hamcon}
\one H \equiv  g_1 \int \one R \; \d^3 x = 0.
\end{equation}
This is a rather weak constraint on the integrated Ricci scalar. Note however, that if we were to abandon projectability then in the current situation we would have the much more restrictive constraint that
\begin{equation}
\one R  = 0.
\end{equation}
We shall have more to say on this point later on in the analysis.

\subsubsection{Supermomentum constraint}

Starting from the extrinsic curvature
\begin{equation}
K_{ij} = - \epsilon {1\over2} \left\{ \dot h_{ij} - \partial_i  n_j - \partial_j n_i \right\}   +  O(\epsilon^2),
\end{equation}
one calculates the conjugate momentum
\begin{equation}
\pi_{ij} =  \epsilon {1\over2} \left\{ \dot h_{ij} - \partial_i  n_j - \partial_j n_i  - (1-\xi) \delta_{ij} \{ \dot h - 2 \vec \partial\cdot \vec n\}  \right\}+  O(\epsilon^2).
\end{equation}
Then the supermomentum constraint becomes
\begin{equation}
\epsilon {1\over2} \partial^i \left\{ \dot h_{ij} - \partial_i  n_j - \partial_j n_i  - (1-\xi) \delta_{ij} \{ \dot h - 2 \vec \partial\cdot \vec n\}  \right\}+  O(\epsilon^2) = 0.
\end{equation}
That is, dropping irrelevant prefactors,
\begin{equation}
\partial^i \left\{ \dot h_{ij} - \partial_i  n_j - \partial_j n_i  - (1-\xi) \delta_{ij} \{ \dot h - 2 \vec \partial\cdot \vec n\}  \right\} = 0.
\end{equation}
It is convenient to rearrange this as
\begin{equation}
\partial^i \left\{ \dot h_{ij}  - (1-\xi) \delta_{ij} \dot h  \right\} 
=
\partial^2 n_j  -  (1-2\xi)  \partial_j (\vec\partial\cdot \vec n). 
\end{equation}
We will soon use an appropriate gauge-fixing to eliminate the RHS, but choose to defer that step for now. Here and below we shall use $\partial^2$ as a convenient shorthand for the flat-space Laplacian $\partial_i \partial^i$.

\subsubsection{Dynamical equation}

We now linearize the dynamical equation. On the LHS
\begin{equation}
{1 \over \sqrt{g}} \; \partial_t ( \sqrt{g} \; \pi^{ij} ) = \epsilon  {1\over2} \partial_t  \left\{ \dot h_{ij} - \partial_i  n_j - \partial_j n_i  - (1-\xi) \delta_{ij} \{ \dot h - 2 \vec \partial\cdot \vec n\}  \right\}+  O(\epsilon^2).
\end{equation}
In counterpoint, on the RHS the only non-trivial contribution comes from linearizing the forcing term $\one F^{ij}$, all other contributions are $O(\epsilon^2)$. Consequently
 \begin{equation}
 {1\over2} \partial_t  \left\{ \dot h_{ij} - \partial_i  n_j - \partial_j n_i  - (1-\xi) \delta_{ij} \{ \dot h - 2 \vec \partial\cdot \vec n\}  \right\}
 =  \one F_{ij}, 
 \end{equation}
which it is convenient to rewrite as
  \begin{equation}
 {1\over2}   \left\{ \ddot h_{ij}  - (1-\xi) \delta_{ij}  \ddot h \right\}
 =     {1\over2}  \left\{  \partial_i  \dot n_j + \partial_j \dot n_i  - 2 (1-\xi) \delta_{ij} ( \vec \partial\cdot \vec{\dot n} )  \right\} + \one F_{ij}.
 \end{equation}
Before presenting the linearization of the forcing term to explicitly obtain $\one F^{ij}$ it is convenient to first discuss gauge fixing.

\subsubsection{Gauge fixing}
\label{sec:gauge}

The projectability condition that we have enforced at the level of the action, {\em i.e.}~the requirement $N=N(t)$, is only part of the gauge freedom usually exploited at the level of the field equations.
The theory under consideration is still invariant under coordinate 
  transformations that preserve the preferred foliation:
\begin{equation}
t\to t+\epsilon \chi^0(t)+ \dots; \qquad x^i \to x^i + \epsilon \chi^i(t,x) + \dots.
\end{equation}
(Note that $\chi^0(t)$ depends only on time, not space).
Then as per the usual (and quite standard) analysis, when working around a flat background
\begin{equation}
g_{ab} \to g_{ab}  + \epsilon \left[ \partial_a \chi_b + \partial_b \chi_a \right] + O(\epsilon^2).
\end{equation}
Extracting the coordinate-transformation-induced changes for the lapse, shift, and 3-metric  (around a flat background)
\begin{eqnarray}
N(t) &\to& N(t) - \epsilon \; \dot{\chi^0}(t) + \dots;
\\
N_i &\to &N_i  +  \epsilon \; \dot\chi_i(t,x) + \dots; 
\\
g_{ij} &\to&  g_{ij} + \epsilon \; [\partial_i \chi_j(t,x) + \partial_j \chi_i(t,x)] + \dots
\end{eqnarray}
and so for the linearized perturbations
\begin{equation*}
n \to n - \dot \chi^0(t);
\end{equation*}
\begin{equation*}
 n_i \to n_i +\dot\chi_i(t,x);
 \end{equation*}
 \begin{equation*}
 h_{ij} \to h_{ij} +\{ \partial_i \chi_j(t,x)+\partial_j \chi_i(t,x) \}; 
 \end{equation*}
 \begin{equation*}
  h \to h +  2\, \partial_i \chi^i(t,x).
\end{equation*}
The so-called \emph{synchronous gauge} consists of choosing $\chi^0(t)$ so that $n(t)\to 0$, and simultaneously choosing $\chi_i(t,x)$ so that $n_i \to 0$.   In this synchronous gauge:
\begin{itemize}
\item
From the supermomentum constraint:
\begin{equation}
\partial^i \left\{ \dot h_{ij}  - (1-\xi) \delta_{ij} \dot h  \right\}
=
0.
\end{equation}
\item
From the dynamical equation:
 \begin{equation}
 {1\over2}   \left\{ \ddot h_{ij}  - (1-\xi) \delta_{ij}  \ddot h \right\}
 =    \one F_{ij}.
 \end{equation}
 \end{itemize}
But even after this synchronous gauge has been adopted there is still
a residual gauge freedom 
\begin{equation}
n \equiv 0;
\qquad
n_i \equiv 0;
 \end{equation}
 \begin{equation}
 h_{ij} \to h_{ij} + \partial_i\bar\chi_j(x) + \partial_j\bar\chi_i(x); 
 \end{equation}
 \begin{equation}
  h \to h +  2\, \partial_i \bar\chi^i(x),
\end{equation}
where $\bar\chi_i(x)$ is \emph{time-independent}. 
This residual gauge freedom is essential to Ho\v{r}ava's prescription for separating the perturbation $h_{ij}$ onto scalar and tensor modes~\cite{Horava}.

\subsubsection{Scalar-tensor decomposition}
One can re-write the supermomentum constraint as
\begin{equation}
 \partial_t \left[ \partial^i \left\{ h_{ij}  - (1-\xi) \delta_{ij}  h  \right\}  \right]=0,
\end{equation}
implying
\begin{equation}
\partial^i \left\{ h_{ij}  - (1-\xi) \delta_{ij}  h  \right\} = K_j(x),
\end{equation}
where $K_j(x)$ is some arbitrary time-independent vector field. Using the residual time-independent coordinate transformations $\bar \chi_i(x)$ we can transform
\begin{equation*}
K_i \to K_i + \partial^2\bar\chi_i - (1-2\xi) \partial_i (\partial_j \bar\chi^j),
\end{equation*}
\begin{equation}
\vec\partial \cdot \vec K \to \vec \partial\cdot \vec K + 2\xi \partial^2  (\vec\partial\cdot \vec{\bar\chi}),
\end{equation}
and suitably choosing $\bar\chi_i(x)$ these residual gauge
transformations can be used to set $K_i(x) \to 0$ as long as $\xi\neq0$ (and this is
effectively what Ho\v{r}ava does~\cite{Horava}). Note that even after setting $K_i(x)\to0$ we are
\emph{still} left with some limited gauge freedom 
\begin{equation}
\bar \chi_i(x) = \partial_i \bar\Psi(x),
\end{equation}
subject to the condition
\begin{equation}
\partial_i \partial^2 \bar \Psi(x)=0,
\end{equation}
as such transformations leave $K_i$ unchanged.
Under this last remaining coordinate freedom
 \begin{equation*}
 h_{ij} \to h_{ij} + 2\partial_i\partial_j\bar \Psi(x); 
 \end{equation*}
 \begin{equation}
  h \to h +  2\, \partial^2 \bar \Psi(x).
\end{equation}  
Unfortunately the exceptional case $\xi=0$ is of most direct relevance to standard general relativity, but let us put that technical issue aside for now. (We will return to this point below.)
If we adopt Ho\v{r}ava's gauge fixing then the super-momentum constraint can effectively be time-integrated to yield
\begin{equation}
\partial^i \left\{ h_{ij}  - (1-\xi) \delta_{ij}  h  \right\} = 0,
\end{equation}
and the only remaining dynamical equation is
 \begin{equation}
 \label{hijdisp}
 {1\over2}   \left\{ \ddot h_{ij}  - (1-\xi) \delta_{ij}  \ddot h \right\}
 =    \one F_{ij}.
 \end{equation}
Taking the trace
\begin{equation}
\label{hdisp}
- \left(1-{3\over2}\; \xi\right) \ddot h = \one F.
\end{equation}
This is the linearized equation of motion for the spin-zero scalar graviton. The potential $\V(g)$, and consequently the forcing term $\one F$, is at this stage generic. (An explicit computation for our particular model will follow shortly.)
To extract the spin-two tensor graviton Ho\v{r}ava then defines (the equivalent of)
\begin{equation}
H_{ij} = h_{ij} - (1-\xi) \delta_{ij} h;  \qquad  
h_{ij} = H_{ij} - {(1-\xi)\over(2-3\xi)} \delta_{ij} H;  \qquad
H = -(2-3\xi) h,
\end{equation}
where $H_{ij}$ is transverse, and furthermore separates out the transverse traceless piece
\begin{equation}
H_{ij} = \widetilde H_{ij} + {1\over2} \left( \delta_{ij} - {\partial_i \partial_j\over\partial^2} \right) H,
\end{equation}
implicitly defining $\widetilde H_{ij}$.
Therefore
\begin{equation}
h_{ij} =   \widetilde H_{ij} + {1\over2} \left( \delta_{ij} - {\partial_i \partial_j\over\partial^2} \right) H - {(1-\xi)\over(2-3\xi)} \delta_{ij} H,
\end{equation}
or
\begin{equation}
h_{ij} =   \widetilde H_{ij} - {(2-3\xi)\over2} \left( \delta_{ij} - {\partial_i \partial_j\over\partial^2} \right) h + {(1-\xi)} \delta_{ij} h.
\end{equation}
{So the dynamical equation~(\ref{hijdisp}) takes the form
 \begin{equation}
 \label{intermediate}
\ddot{ \widetilde H}_{ij} 
 =        2 \; \one F_{ij} -  {1\over2} \left( \delta_{ij} - {\partial_i \partial_j\over\partial^2} \right) \ddot H
 =        2 \; \one F_{ij} +  {(2-3\xi)\over2} \left( \delta_{ij} - {\partial_i \partial_j\over\partial^2} \right) \ddot h.
 \end{equation}
But in view of the dispersion relation for $h$ already derived in equation~(\ref{hdisp}), this can easily be
rewritten as}
\begin{equation}
\ddot{ \widetilde H}_{ij} 
 =        2 \; \one F_{ij} - \left( \delta_{ij} - {\partial_i \partial_j\over\partial^2} \right) \one F,
 \end{equation}
 so
 \begin{equation}
\ddot{ \widetilde H}_{ij}  = 2\left( \one F_{ij} - {1\over2}  \delta_{ij} \;\one F \right) +  {\partial_i \partial_j\over\partial^2} \; [\,\one F\,].
\end{equation}
This is the linearized equation of motion for the spin-two tensor graviton. The potential $\V(g)$, and consequently the forcing term $\one F_{ij}$, is at this stage generic. (An explicit computation for our particular model will follow shortly.)
It is easy to check that the RHS of the above is both transverse and traceless, as is required for consistency.
The analysis of this section has now decomposed the metric perturbation $h_{ij}$ onto a spin-0 scalar, $h$, and a spin-2 tensor $ {\widetilde H}_{ij}$, with appropriate equations of motion for each.  The only remaining task is to calculate the linearized quantities $\one F_{ij}$ and $ \one F$ for our particular model, using the gauge-fixing to simplify terms as much as possible.

\subsubsection{The exceptional $\xi=0$ case}

In the exceptional case $\xi=0$ the supermomentum constraint after going to synchronous gauge (and before any residual gauge fixing) is 
\begin{equation}
 \partial_t \left[ \partial^i \left\{ h_{ij}  - \delta_{ij}  h  \right\}  \right]=0,
\end{equation}
implying
\begin{equation}
\partial^i \left\{ h_{ij}  -  \delta_{ij}  h  \right\} = K_j(x),
\end{equation}
where again $K_j(x)$ is some arbitrary time-independent vector field. Using the residual time-independent coordinate transformations $\bar \chi_i(x)$ we can transform
\begin{equation}
K_i \to K_i + \partial^2\bar\chi_i -  \partial_i (\partial_j \bar\chi^j) ,
\end{equation}
implying
\begin{equation}
K_{[i,j]} \to K_{[i,j]} + \partial^2\bar\chi_{[i,j]};
\qquad\qquad
\vec\partial \cdot \vec K \to \vec \partial\cdot \vec K.
\end{equation}
So in this exceptional case we cannot eliminate $K_i$ completely, though we \emph{can} eliminate $K_{[i,j]}$. (To do this choose $\partial^2 \bar\chi_i$ to be any of the ``vector potentials'' leading to the ``magnetic field" $-K_{[i,j]}$. This will specify $\bar\chi_i(x)$ up to a gauge transformation $\partial_i \bar \Psi(x)$.)  Thus, after this next step of gauge fixing, there is some time independent scalar $\Phi(x)$ such that $K_i(x) = \partial_i \Phi(x)$ and so the (time-integrated) supermomentum constraint becomes
\begin{equation}
\partial^i \left\{ h_{ij}  -  \delta_{ij}  (h +\Phi) \right\} = 0,
\end{equation}
and we \emph{still} have available two residual gauge transformations
coming from the quantity $\bar \Psi(x)$:
\begin{equation}
\label{xi0gauge}
 h_{ij} \to h_{ij} - 2  \partial_i\partial_j \bar \Psi(x); 
\qquad  h \to h - 2 \; \partial^2 \bar \Psi(x).
\end{equation}
The dynamical equation is now
 \begin{equation}
 \label{hijdisp:b}
 {1\over2}   \left\{ \ddot h_{ij}  -  \delta_{ij}  \ddot h \right\}
 =    \one F_{ij}.
 \end{equation}
Taking the trace
\begin{equation}
\label{hdisp:b}
- \ddot h = \one F.
\end{equation}
To obtain the transverse-traceless mode suitable for this situation we first define
\begin{equation}
H_{ij} = h_{ij} - \delta_{ij} (h+\Phi);  \qquad  
h_{ij} = H_{ij} - {H+\Phi\over2} \delta_{ij};  \qquad
H = -2 h- 3 \Phi,
\end{equation}
where $H_{ij}$ is transverse, and again separate out the transverse traceless piece
\begin{equation}
H_{ij} = \widetilde H_{ij} + {1\over2} \left( \delta_{ij} - {\partial_i \partial_j\over\partial^2} \right) H,
\end{equation}
implicitly defining $\widetilde H_{ij}$.
Therefore
\begin{equation}
h_{ij} =   \widetilde H_{ij} + {1\over2} \left( \delta_{ij} - {\partial_i \partial_j\over\partial^2} \right) H - {H+\Phi\over2}\; \delta_{ij},
\end{equation}
or
\begin{equation}
h_{ij} =   \widetilde H_{ij} - {1\over2} \left( \delta_{ij} - {\partial_i \partial_j\over\partial^2} \right) (2h+3\Phi) + \delta_{ij} (h+\Phi).
\end{equation}
So (remembering that $\Phi(x)$ is time independent) the dynamical equation~(\ref{hijdisp:b}) now takes the \emph{same} form as for $\xi\neq 0$.
 \begin{equation}
 \label{intermediate:b}
\ddot{ \widetilde H}_{ij} 
 =        2 \; \one F_{ij} -  {1\over2} \left( \delta_{ij} - {\partial_i \partial_j\over\partial^2} \right) \ddot H
 =        2 \; \one F_{ij} +  \left( \delta_{ij} - {\partial_i \partial_j\over\partial^2} \right) \ddot h.
 \end{equation}
But in view of the dispersion relation for $h$ already derived in equation~(\ref{hdisp:b}), this can easily be
rewritten as
\begin{equation}
\ddot{ \widetilde H}_{ij} 
 =        2 \; \one F_{ij} - \left( \delta_{ij} - {\partial_i \partial_j\over\partial^2} \right) \one F = 
\ddot{ \widetilde H}_{ij}  = 2\left( \one F_{ij} - {1\over2}  \delta_{ij} \;\one F \right) +  {\partial_i \partial_j\over\partial^2} \; [\,\one F\,].
\end{equation}
So despite  extra technical complications for $\xi=0$, the ultimate set of differential equations looks very similar --- though we shall still see some subtleties arise after explicit linearization of the forcing term.

\subsection{Linearizing the forcing term}

We can now return to the linearization of the forcing term $F_{ij} \to \one F_{ij}$. Using the general form for the forcing term without yet imposing any gauge conditions and specializing to a Minkowski background
\begin{eqnarray}
 \one (F_1)_{ij} &=&  \one R_{ij} - {1\over2} (\one R) \delta_{ij};
 \\
  \one (F_2)_{ij} &=& -2 [\partial_i \partial_j - \delta_{ij} \partial^2] (\one R);
  \\
 \one (F_3)_{ij} &=&  \partial^2 (\one R_{ij}) +{1\over2} \delta_{ij} \partial^2 (\one R) - \partial_i \partial_j (\one R); 
 \\
  \one (F_4)_{ij} &=&  0;
   \\
  \one (F_5)_{ij}  &=&  0;
   \\
  \one (F_6)_{ij} &=&  0;
  \\
   \one (F_7)_{ij} &=&  2  [\partial_i \partial_j - \delta_{ij} \partial^2] \partial^2 (\one R);
   \\
  \one (F_8)_{ij} &=&  \partial^4 (\one R_{ij})  -  \partial^2 \left[ \partial_i \partial_j - {1\over2}  \delta_{ij}   \partial^2  \right](\one R).
\end{eqnarray}
Similarly for the trace
\begin{eqnarray}
 \one (F_1)&=&   - {1\over2} (\one R);
 \\
  \one (F_2)&=& 4 \partial^2 (\one R);
  \\
 \one (F_3) &=& {3\over2} \partial^2 (\one R); 
 \\
  \one (F_4) &=&  0;
   \\
  \one (F_5)  &=&  0;
   \\
  \one (F_6) &=&  0;
  \\
   \one (F_7) &=&  -4 \partial^4 (\one R);
   \\
  \one (F_8) &=&  {3\over2} \partial^4 (\one R).
\end{eqnarray}
But calculating $\one R_{ij}$ and $\one R$ is simple. Before any gauge fixing
\begin{equation*}
\one R_{ij} = -{1\over2} (\partial^2 h_{ij} + \partial_i \partial_j h - h_{mi,mj} - h_{mj,mi} ),
\end{equation*}
\begin{equation}
\one R = - (\partial^2 h - h_{mn,mn} ).
\end{equation}
Applying Ho\v{r}ava's gauge condition ($\xi\neq0$, plus synchronous gauge, plus in addition the residual gauge fixing), 
\begin{eqnarray*}
\one R_{ij} &\to&  
-{1\over2} (\partial^2 h_{ij} - \partial_i \partial_j h)  - \xi (\partial_i \partial_j h)
=  
-{1\over2} \left[\partial^2 {\widetilde H}_{ij} +{\xi\over2} (\delta_{ij} \partial^2 + \partial_i \partial_j) h \right]; \qquad
\end{eqnarray*}
\begin{equation}
\label{xineq0R}
\one R  \to -\xi \partial^2 h.
\end{equation}
This allows us to assert
\begin{eqnarray}
 \one (F_1)_{ij} &=&  -{1\over2} \partial^2 {\widetilde H}_{ij} + \xi \hbox{ (derivative terms acting on) } h;
 \\
  \one (F_2)_{ij} &=&  \xi \hbox{ (derivative terms acting on) } h;
  \\
 \one (F_3)_{ij} &=&    -{1\over4} \partial^4 {\widetilde H}_{ij} + \xi \hbox{ (derivative terms acting on) } h;
  \\
  \one (F_4)_{ij} &=&  0;
   \\
  \one (F_5)_{ij}  &=&  0;
   \\
  \one (F_6)_{ij} &=&  0;
  \\
   \one (F_7)_{ij} &=&  \xi \hbox{ (derivative terms acting on) } h;
   \\
  \one (F_8)_{ij} &=&   -{1\over2} \partial^6 {\widetilde H}_{ij} + \xi \hbox{ (derivative terms acting on) } h;
\end{eqnarray}
For the trace we have
\begin{eqnarray}
 \one (F_1)&=&    {1\over2}\xi\partial^2 h ;
 \\
  \one (F_2)&=&  {-4}\xi\partial^4 h ;
  \\
 \one (F_3) &=&  -{3\over2}\xi\partial^4 h ;
 \\
  \one (F_4) &=&  0;
   \\
  \one (F_5)  &=&  0;
   \\
  \one (F_6) &=&  0;
  \\
   \one (F_7) &=&  {4}\xi\partial^6 h ;
   \\
  \one (F_8) &=&  -{3\over2}\xi\partial^6 h ;
\end{eqnarray}
In contrast, in the exceptional situation $\xi=0$ we obtain
\begin{eqnarray*}
\!\!\!
\one R_{ij} &\to&  -{1\over2} (\partial^2 h_{ij} - \partial_i \partial_j h)  +\partial_i \partial_j \Phi(x)
=
 -{1\over2} \left[\partial^2 {\widetilde H}_{ij} -{1\over2} (\delta_{ij} \partial^2 + \partial_i \partial_j) \Phi(x) \right];
 \qquad\quad
\end{eqnarray*}
\begin{equation}
\one R  \to \partial^2 \Phi(x).
\end{equation}
Note in particular that in the $\xi=0$ case we see that $\one R$ is ``frozen in''; because of the interplay between supermomentum constraints and gauge fixing all the time-dependence in $\one R$ has been eliminated.

\subsection{Spin-0 scalar graviton}
It is most efficient to separate the discussion of the scalar graviton into three separate cases.
\subsubsection{General kinetic term ($\xi\neq0$):}
Collecting terms from the above, for the spin-0 scalar graviton, when $\xi\neq 0$  the linearized equation of motion is simply
\begin{equation}
\label{xineq0scalar}
\left(1-{3\over2}\, \xi\right)   \; \ddot h = - \xi \left\{ {1\over2} g_1 \partial^2 +  \left(-4 g_2 - {3\over2} g_3 \right) \zeta^{-2} \partial^4 + \left(4 g_7 -{3\over2} g_8\right) \zeta^{-4} \partial^6 \right\} h.
 \end{equation}
This is a sixth-order trans-Bogoliubov dispersion relation~\cite{LSB-regulator} for the scalar mode $h$.
Note that only \emph{some} of the potential couplings contribute here, ($g_1$, $g_2$, $g_3$, $g_7$, $g_8$), and it is a \emph{different} set of couplings from what we shall soon see is relevant for the tensor graviton ($g_1$, $g_3$, $g_8$), or is relevant for FLRW cosmology ($g_0$---$g_6$).   Combining the Hamiltonian constraint in equation~(\ref{hamcon}) with equation~(\ref{xineq0R}) we can at least deduce that 
\begin{equation}
\int \partial^2 h(t,x) \; \d^3 x=0.
\end{equation}
In combination with the equation of motion for the spin-0 excitation, see equation (\ref{xineq0scalar}), this implies
\begin{equation}
\partial_t^2  \int h \; \d^3 x =0.
\end{equation}
But then  $\int h \; \d^3 x$ is linear in time:
\begin{equation}
\label{constr}
 \int h(t,x) \; \d^3 x =   \int h(0,x) \; \d^3 x+   t\, \int \dot h(0,x) \; \d^3 x.
 \end{equation}
If we now use the
 very last bit of coordinate freedom, (the function $\bar \Psi(x)$),
 we can only set $\int h(0,x) \; \d^3 x \to 0$, but 
 we cannot choose to fix  $\int \dot h(0,x) \; \d^3 x$.  So the total volume
 cannot be fixed:
\begin{equation}
\label{constr2}
\int h(t,x) \; \d^3 x \to   t\; \int \dot h(0,x) \; \d^3 x.
\end{equation}
Thus $h$ is given by the solution of equation~(\ref{xineq0scalar}) subject
 to the constraint equation~(\ref{constr2}). 

\subsubsection{Specific general relativistic kinetic term ($\xi=0$):}
For $\xi=0$ we need a separate special-case analysis. In view of our computations above, one merely has to make the formal replacements $\xi h \to -\Phi(x)$  followed by $\xi \to 0$ to obtain
\begin{equation}
\label{xi0scalar}
 \ddot h = \left\{ {1\over2} g_1 \partial^2 +  \left(-4 g_2 - {3\over2} g_3 \right) \zeta^{-2} \partial^4 + \left(4 g_7 -{3\over2} g_8\right) \zeta^{-4} \partial^6 \right\} \Phi(x).
 \end{equation}
Note that this is \emph{not} a ``wave equation''. The quantity $\Phi(x)$ is by construction time-independent, so this says that at each spatial point $x$ the scalar mode $h(t,x)$ is undergoing ``constant acceleration''. Indeed equation~(\ref{xi0scalar}) can readily be solved to give
\begin{equation}
h(t,x) = h(0,x) +\dot h(0,x)\, t + {1\over2} \ddot h(0,x) t^2. 
\end{equation}
According to equation~(\ref{xi0gauge}), we still have just enough remaining 
  gauge freedom to set $h(0,x)\rightarrow 0$ and get
\begin{equation}
h(t,x) = \dot h(0,x)\, t +{1\over2} \ddot h(0,x) t^2
\end{equation}
(the function $\bar\Psi(x)$ can now be chosen arbitrarily). Note that $\ddot h(0,x)$ is just the right hand side of equation~(\ref{xi0scalar}). Using the linearized Hamiltonian constraint in equation~(\ref{hamcon}) as above now gives
\begin{equation} 
\int \partial^2 \Phi(x) \; \d^3 x=0,
\end{equation}
and so
\begin{equation} 
\int  \ddot h(t,x) \; \d^3 x= \int  \ddot h(0,x) \; \d^3 x=0,
\end{equation}
which is the only constraint on the functional form of $\ddot
h(0,x)$. 
Again, the total volume of space is not fixed.

\subsubsection{Non-projectable case:}

Let us briefly examine now what would happen if we would decide to relax the projectability condition. In this case the Hamiltonian constraint would become a super-Hamiltonian constraint $\one R = 0$, which would imply, for the $\xi\neq 0$ case $\partial^2 h(t,x)=0$, and for the $\xi=0$ case $\partial^2 \Phi(x)=0$.
\begin{itemize}
\item 
For  $\xi\neq 0$ the constraint $\partial^2 h=0$, together with a boundedness condition on $h(t,x)$ and suitable fall-off conditions at spatial infinity, then implies 
\begin{equation}
h(t,x) = 0,
\end{equation}
so in this particular case the \emph{linearized} scalar mode can be eliminated.\footnote{A similar result has subsequently been derived in a cosmological setting in reference~\cite{Brandenberger2}.} 
\item
In contrast, for $\xi=0$ the equation of motion collapses to $\ddot h =0$ for all $x$, and so 
\begin{equation}
h(t,x) = h(0,x) +\dot h(0,x)\, t. 
\end{equation}
So in the $\xi=0$ non-projectable case there is not enough
residual gauge freedom to set $h$ to zero.
We appear to be left with an undesirable linear expansion mode. 
\end{itemize}
Overall, the behaviour of the spin-zero mode is certainly disturbing and definitely deserves fuller investigation. (Note in particular that in a different context the authors of reference~\cite{Cai3} have also encountered difficulties with the spin-zero mode).\footnote{The fact that for $\xi \neq 0$ the scalar mode can be eliminated at the level of linear perturbations does not necessarily mean it is absent altogether. Subsequent to the original appearance of this article, several papers have discussed this issue --- as well as further difficulties that arise beyond the linearized level \cite{Problems,Li-Pang,Blas}. Note also that the linear expansion mode found for the $\xi=0$ non-projectable case need not be physical. Indeed, it was again subsequently shown in \cite{Wang:2009yz} that it is essentially a gauge artifact.}

\subsection{Spin-2 tensor graviton}

Extracting the equation for the spin-2 tensor graviton requires a little more work. From the results assembled for $\one F_{ij}$ above, we can assert
\begin{equation}
\ddot{ \widetilde H}_{ij}   =  -\left[  g_1  \partial^2   + g_3  \zeta^{-2} \partial^4   + g _8  \zeta^{-4} \partial^6 \right] \widetilde H_{ij} + \xi X_{ij}.
\end{equation}
Here $X_{ij}$ is a tensor that is linear in the scalar $h$, independent of $\xi$, and whose tensor structure arises solely from partial derivatives and the Kronecker delta. Furthermore by construction $X_{ij}$ is both transverse and traceless. 
This tells us
\begin{equation}
X_{ij} =    \{  a(\partial^{2n}) \partial^2 \delta_{ij} + b (\partial^{2n}) \partial_i \partial_j \} h.
\end{equation}
But tracelessness implies $3a+b=0$ while transversality implies $a+b=0$.  Therefore $a=b=0$ and the tensor $X_{ij}=0$. So the spin-2 tensor graviton satisfies
\begin{equation}
\ddot{ \widetilde H}_{ij}   =  -\left[  g_1  \partial^2   + g_3  \zeta^{-2} \partial^4   + g _8  \zeta^{-4} \partial^6 \right] \widetilde H_{ij}.
\end{equation}
This is again a sixth-order trans-Bogoliubov dispersion relation~\cite{LSB-regulator}, now sensitive to a different combination of the coupling constants.
This dispersion relation does not depend on $\xi$, and careful analysis of the special case $\xi=0$ shows this equation for the tensor mode continues to hold even for that otherwise exceptional case.

\subsection{Low-momentum and high-momentum limits}
In the low-momentum limit the spin-2 graviton is naively seen to have both phase velocity and group velocity $c_\mathrm{spin\,2}^2 \to - g_1$. But we had already noted that in going to physical normalization for the Einstein--Hilbert term we need to scale space and time so that $g_K\to +1$ and $g_1 \to -1$. So, as expected, the physical phase and group velocities of the low-momentim spin 2 graviton are simply $c_\mathrm{spin\,2}^2 \to 1$. (This is very strongly reminiscent of the way in which the Bogoliubov dispersion relation leads in the low-momentum limit to a phononic branch with finite speed of sound. See, for example,~\cite{Living, broken}.) In short, the behaviour of the tensor mode is physically reasonable.

In contrast, the low-momentum phase velocity (and group velocity) of the spin-0 graviton is seen to be
\begin{equation}
c_\mathrm{spin\,0}^2 =  - {\xi \; g_1\over 2-3\xi} \to   {\xi \; \over 2-3\xi}.
\end{equation}
So at low momentum this spin-0 mode is propagating (hyperbolic) for $\xi\in(0, 2/3)$, but is non-propagating (elliptic) for  $\xi<0$ and $\xi>2/3$. 
Elliptic modes are potentially dangerous in that they correspond to imaginary frequencies and can lead to exponential instabilities, (see for instance~\cite{White1, White2}).
Note that whether this mode is propagating (hyperbolic) or non-propagating (elliptic) can now depend on the momentum. For examples of similar phenomena in the ``analogue spacetime'' framework, see~\cite{White1, White2}. The situation is actually worse than this: In Ho\v{r}ava's parameterization
\begin{equation}
  {\xi \over 2-3\xi} = - {1-\lambda\over  1-3\lambda},
\end{equation}
so a propagating scalar mode corresponds to a negative kinetic energy \emph{ghost mode}. (See especially (4.56) of~\cite{Horava2}.) Since the presence of a scalar ghost depends only on the kinetic term $\T(K)$, this feature is likely to be generic to all Ho\v{r}ava-like models. (In fact, if one takes seriously the supposed renormalization group running of Ho\v{r}ava's original gravity model~\cite{Horava} from a $\lambda=1/3$ [$\xi=2/3$] conformal coupling in the ultraviolet to a $\lambda=1$ [$\xi=0$] Einstein-Hilbert coupling in the infrared, then this evolution has to take one through the parameter region where the scalar mode kinetic energy is negative.) These observations again strongly suggest that it is desirable to eliminate the scalar mode if at all possible.

Turning to high momentum,  the dispersion relation in that regime can be either sub-linear or super-linear depending on the signs of the appropriate couplings ---  in the ``analogue spacetime'' framework~\cite{Living} such dispersion relations are referred to as ``subluminal'' or ``superluminal'' respectively. (In that framework such modified dispersion relations are typically used as ways of probing the physics of Hawking radiation or cosmological particle production~\cite{Liberati2, Cosmo1, Cosmo2}, but in the present context the (momentum)$^4$ and (momentum)$^6$ terms are to be viewed as fundamental physics.)

\section{Application to FLRW cosmology}
In a cosmological setting the forcing terms $F_{ij}$ simplify tremendously, and provide one with a simple modification of standard FLRW cosmology. (As we shall soon see, the Friedmann equations pick up a few extra terms from the higher-derivative spatial curvature terms in $\V(g)$, and the kinetic part of the Friedmann equations is slightly modified by the $\xi$ term in $\T(K)$.) The 3+1 line element is ($k\in\{-1,0,+1\}$)
\begin{equation}
\d s^2 = - c^2 \d t^2 + a(t)^2 \left\{ {\d r^2\over 1- k r^2} + r^2 \left(\d\theta^2 +\sin^2\theta\;\d\phi^2\right)\right\}.
\end{equation}
\subsection{First Friedmann equation}
The extrinsic curvature of the spatial slices is
\begin{equation}
K_{ij} =  -{\dot a(t) \over a(t)}\; g_{ij};  \qquad  K = -{3 \dot a(t) \over a(t)},
\end{equation}
whence
\begin{equation}
\T(K) \to -(6-9\xi) \; {\dot a^2\over a^2}.
\end{equation}
Since the spatial slices are (by the definition of FLRW spacetime) constant-curvature hypersurfaces we have
\begin{equation}
R_{ij} =  {2 k \over a^2(t)}\; g_{ij};  \qquad  R = {6 k \over a^2(t)}.
\end{equation}
So in any FLRW spacetime
\begin{eqnarray}
\V(g)  &\to&   \V(a) = g_0\, \zeta^6 +  {6 g_1 k \, \zeta^4 \over a^2} + {12(3g_2+g_3) \zeta^2 k^2\over a^4} 
+ {24(9 g_4 + 3g_5+ g_6) k\over a^6}. \qquad
\end{eqnarray}
(Note that the $g_7$ and $g_8$ terms drop out due to the translation invariance in the spatial slices, and that because of its definition $k^{2n+1} = k$ and $k^{2n}=k^2$. ) 
In an idealized \emph{exact} FLRW spacetime, because of the spatial homogeneity, the spatial integral can be dropped from the Hamiltonian constraint, which  now simplifies to 
\begin{equation}
\label{1stfried}
\T(g)+\V(g)=0,
\end{equation}
leading to the first (vacuum) Friedmann equation (currently in $Z=1$ units):
\begin{equation}
\left(1-{3\over2}\xi\right) \; {\dot a^2\over a^2} =  {\V(a)\over 6}.
\end{equation}
This version of the first Friedmann equation will hold in any Ho\v{r}ava-like model in that it depends only on the symmetries of FLRW spacetime and the general features of the model, such as the structure of the kinetric term $\T(K)$, and is independent of the specific details of the potential $\V(a)$.
For our specific variant of Ho\v{r}ava's model:
\begin{equation}
\left(1-{3\over2}\xi\right) \; {\dot a^2\over a^2} =  {g_0\, \zeta^6\over6} +  { g_1 k \, \zeta^4 \over a^2} + {2(3g_2+g_3) \zeta^2 k^2\over a^4} 
+ {4(9 g_4 + 3g_5+ g_6) k\over a^6}.
\end{equation}
Going over to $c=1$ units, and rearranging somewhat
\begin{equation}
\left(1-{3\over2}\xi\right) \; {\dot a^2\over a^2} +  {k\over a^2} =  {\Lambda\over3}  + {2(3g_2+g_3) \zeta^{-2} k^2\over a^4} 
+ {4(9 g_4 + 3g_5+ g_6) \zeta^{-4} k\over a^6}.
\end{equation}
So even in the absence of explicit matter fields, the $g_2$ and $g_3$ terms can mimic the effects of radiation pressure (``dark radiation''), of either sign depending on the sign of the relevant coupling constants. (See also~\cite{cosmology1}.) Similarly, the $g_4$, $g_5$, and $g_6$ terms can mimic stiff matter of either sign depending on the sign of spatial curvature and the sign of the relevant coupling constants. (Recall that for or stiff matter  $\rho=p$ and so $\rho_\mathrm{stiff} \propto 1/a^6$.) Note that the stiff-matter mimicking term is absent in Ho\v{r}ava's original model~\cite{cosmology1}. This is due to the fact that the the only $[\kappa]^6$ term in Ho\v{r}ava's original model is (Cotton)$^2$, and the Cotton tensor automatically vanishes for the constant curvature spatial slices of a FLRW spacetime. 

We now add a cosmological fluid through a purely pragmatic and phenomenological approach: 
Working within the cosmological hydrodynamical approximation we approximate the cosmological stress-energy tensor by two quantities, the density $\rho$ and pressure $p$, and simply add them to the vacuum equations by demanding the correct limit as one approaches classical general relativity.  (More fundamentally, we could attempt to derive the matter contributions from an action principle, but for the present purposes that would be overkill. A model for the matter sector based on a simple scalar field is outlined in~\cite{cosmology1,cosmology2}, and the resulting Friedmann equations [insofar as there is an overlap] are compatible with our purely pragmatic approach.) Under these assumptions the first Friedmann equation becomes
\begin{equation}
\left(1-{3\over2}\xi\right) \; {\dot a^2\over a^2} +  {k\over a^2} =  {\Lambda\over3} + {2(3g_2+g_3) \zeta^{-2} k^2\over a^4} 
+ {4(9 g_4 + 3g_5+ g_6) \zeta^{-4} k\over a^6} + {\rho\; \zeta^{-2}\over 6},
\end{equation}
or equivalently
\begin{equation}
\label{1stfried}
\left(1-{3\over2}\xi\right) \; {\dot a^2\over a^2} +  {k\over a^2} =  {\Lambda\over3} + {2(3g_2+g_3) \zeta^{-2} k^2\over a^4} 
+ {4(9 g_4 + 3g_5+ g_6) \zeta^{-4} k\over a^6} + {8\pi G_\mathrm{N} \rho\over 3}.
\end{equation}
We this see a controlled deviation from standard cosmology, with the deviations from standard cosmology being governed (effectively) by three parameters: $\xi$,  $(3g_2+g_3)$, and $(9 g_4 + 3g_5+ g_6)$. For a generic Ho\v{r}ava-like model coupled to matter we would have
\begin{equation}
\left(1-{3\over2}\xi\right) \; {\dot a^2\over a^2}  =  {\V(a)\over 6}  + {8\pi G_\mathrm{N} \rho\over 3}.
\end{equation}
 It is worth pointing out that the presence of the second and third term in the right hand side of the first Friedmann equation~(\ref{1stfried}) could potentially lead to bouncing solutions for suitable values of the parameters. However, the behaviour of matter contribution close to the bounce would be critical and further investigation is needed (see also~\cite{Brandenberger}).  In this regard it is perhaps worthwhile recalling that a bounce requires overall violation of the strong energy condition (SEC) in its immediate vicinity~\cite{Molina, Cattoen}.
 
Before going further, a remark is appropriate concerning the passage from the Hamiltonian constraint to equation~(\ref{1stfried}): Dropping the integral over space depends crucially on exact homogeneity (hence the emphasis on the word ``exact'' there). Therefore, this step will not be possible in more general cosmological contexts, where the universe is not exactly FLRW, but can instead be interpreted as a large number of almost FLRW sub-regions. In this case one would have to develop a more complete treatment.\footnote{Subsequent to the original appearance of this article, Mukohyama~\cite{dark-dust}  has advocated the viewpoint that, in such a setting, the integrated Hamiltonian constraint can be re-interpreted as the ordinary Friedmann equation with the addition of ``dark dust'' that is comoving with the preferred foliation.}
 
\subsection{Second Friedmann equation}
The second Friedmann equation comes from the dynamical equation for the canonical momentum $\pi^{ij}$.  For the conjugate momentum in a FLRW universe 
\begin{equation}
\pi^{ij} \to -(2 - 3 \xi) \; {\dot a\over a} \; g^{ij},
\end{equation}
and so equation~(\ref{piequation}) takes the form
\begin{equation}
-a^{-3} \partial_t \left( a^3 \;  (2 - 3 \xi) \; {\dot a\over a} \; g^{ij} \right)  =  2(2 - 3 \xi)\; {\dot a^2\over a^2} \; g^{ij} + {1\over2} \T(K) \;g^{ij} + F^{ij}.
\end{equation}
This leads to
\begin{eqnarray}
- \left(1 - {3\over2} \xi\right) \; {\ddot a\over a}   \; g^{ij}  &=&  + {1\over2}\left(1 - {3\over2} \xi\right) \;  {\dot a^2\over a^2}    \; g^{ij}  + {F^{ij}\over2}.
\end{eqnarray}
Taking the trace
\begin{eqnarray}
- \left(1 - {3\over2} \xi\right) \; {\ddot a\over a}    &=&  + {1\over2}\left(1 - {3\over2} \xi\right) \;  {\dot a^2\over a^2}   + {g_{ij} F^{ij}\over6}.
\end{eqnarray}
But from the definition of the forcing term $F_{ij}$, its trace satisfies 
\begin{equation}
g_{ij} \; F^{ij} =  g_{ij} \; {1\over \sqrt{g}}  \;  {\delta S_\V \over\delta g_{ij} } =  -{1\over 2 a^2} {\d[\V(a)\, a^3]\over \d a}. 
\end{equation}
So the dynamical equation for the conjugate momentum becomes ($Z=1$ units, still for vacuum)
\begin{eqnarray}
- \left(1 - {3\over2} \xi\right) \; {\ddot a\over a}   &=&  + {1\over2}\left(1 - {3\over2} \xi\right)  {\dot a^2\over a^2}   
- {1\over12 a^2} {\d[\V(a)\, a^3]\over\d a}.
\end{eqnarray}
In physical ($c\to1$) units the only change is
\begin{eqnarray}
- \left(1 - {3\over2} \xi\right) \; {\ddot a\over a}   &=&  + {1\over2}\left(1 - {3\over2} \xi\right)  {\dot a^2\over a^2}   
- \zeta^{-4} {1\over12 a^2} {\d[\V(a)\, a^3]\over\d a}.
\end{eqnarray}
Adding a (phenomenological) cosmological fluid simply adds a pressure contribution
\begin{eqnarray}
- \left(1 - {3\over2} \xi\right) \; {\ddot a\over a}   &=&  + {1\over2}\left(1 - {3\over2} \xi\right)  {\dot a^2\over a^2}   
- \zeta^{-4} {1\over12 a^2} {\d[\V(a)\, a^3]\over\d a}
+ 4\pi G_N p.\qquad
\end{eqnarray}
We again see a controlled deviation from standard cosmology, with the deviations from standard cosmology being governed by $\xi$ and by $\V(a)$ which is at this stage of the calculation arbitrary.

For our specific variant of Ho\v{r}ava's model the explicit forcing terms $F_{ij}$ also simplify tremendously in FLRW cosmologies. 
 In particular
\begin{eqnarray}
(F_0)_{ij} &\to& -{1\over2} g_{ij}.\\
(F_1)_{ij} &\to& -{k\over a^2(t)} \; g_{ij} .\\
(F_2)_{ij} &\to&  {6 k^2\over a^4(t)} \; g_{ij}.\\
(F_3)_{ij} &\to&   {2 k^2\over a^4(t)} \; g_{ij}. \\
(F_4)_{ij} &\to&   {108 k\over a^6(t)} \; g_{ij}. \\
(F_5)_{ij} &\to&  {36 k\over a^6(t)} \; g_{ij}. \\
(F_6)_{ij}  &\to& {12 k\over a^6(t)} \; g_{ij}.\\
(F_7)_{ij} &\to& 0.\\
(F_8)_{ij}  &\to& 0.
\end{eqnarray}
That is, inserting suitable dimensional factors
\begin{equation}
F_{ij} = \left\{  -{g_0\,\zeta^6\over2} - {g_1 k\, \zeta^4 \over a^2} +{2(3g_2+g_3) k^2 \, \zeta^2\over a^4}  +{12(9 g_4 + 3 g_5 + g_6)  k\over a^6}  \right\} \; g_{ij}.
\end{equation}
It is then easy to explicitly verify that for our particular variant of Ho\v{r}ava's model
\begin{equation}
F_{ij} = \left\{ - {1\over2}\V(a) - {a\over6} {\d\V(a)\over\d a}  \right\} \; g_{ij}  = -{1\over6 a^2} {\d[\V(a) a^3] \over\d a}  \; g_{ij}.
\end{equation}
The deviations from standard cosmology in our model are governed (effectively) by three parameters: $\xi$,  and the compound parameters $(3g_2+g_3)$ and $(9 g_4 + 3g_5+ g_6)$ hiding inside the potential $\V(a)$.

\subsection{Third Friedmann equation}
By eliminating the $\dot a^2/a^2$ term between the first and second Friedmann equations one sees
\begin{eqnarray}
- \left(1 - {3\over2} \xi\right) \; {\ddot a\over a}   &=&  - \left\{  {1\over6}\V(a) + {a\over12} {\d\V(a)\over\d a}  \right\} =  -{1\over12 a} {\d[\V(a) a^2] \over\d a}  
\end{eqnarray}
Going to $c\to1$ units, and adding (phenomenological) cosmological pressure and density, 
\begin{eqnarray}
- \left(1 - {3\over2} \xi\right) \; {\ddot a\over a}   &=&    -{1\over12 a} {\d[\V(a) a^2] \over\d a}  +{4\pi G_\mathrm{N}\over3} (\rho+3p),
\end{eqnarray}
with $\V(a)$ still being arbitrary at this stage. 
In summary, most of FLRW cosmology survives with only minimal changes. There is a small change in the ``kinetic'' part of the Friedmann equations, easily dealt with by inserting a factor of $(1-{3\over2}\xi)$, and in the ``potential'' the higher-spatial-curvature terms mimic various forms of matter. In our specific model the $\V(a)$ terms mimic radiation pressure (``dark radiation'') and stiff matter.

\section{Discussion}
The specific extension of Ho\v{r}ava's model that we have outlined above
so far only considers pure gravity (and a phenomenologically introduced notion of cosmological fluid). It is however a very definite proposal with a small number of adjustable parameters, (many fewer adjustable parameters than the standard model of particle physics), making it worthwhile to put in the additional effort to develop precision tests that would confront this model with experimental and observational bounds.  

At this stage we have investigated the graviton propagators (and weak-field gravitational waves) around flat Minkowski space, both for the spin-0 scalar and spin-2 tensor gravitons, and demonstrated how they depend on the specific terms in the potential $\V(g)$. The presence and behaviour of the spin-0 mode appear to be worrying,  and definitely require further investigation.\footnote{After the original upload of this this article we became aware of reference~\cite{Problems}, which also seems to suggest that serious problems are introduced by the scalar mode.} We have also investigated FLRW cosmologies, and seen how the Freidmann equations are modified for generic $\V(g)$.  We have seen that the Friedmann equations and the graviton propagators are sensitive to different subsets of the coupling constants and so can in principle probe different parts of the physics. 

The most significant theoretical restriction we have retained is Ho\v{r}ava's ``projectability'' constraint. Our analysis showed that there are a number of reasons why it might be useful to relax this constraint. Based on the results of this article we can make some suggestions as to what might happen in the absence of a projectability constraint:
\begin{itemize}
\item One would have to modify the use of the integration by parts argument when discarding surface terms. Some of the terms we discarded when constructing our list of all possible terms appearing in $\V(g)$  would have to be retained.  (So $\V(g)$  would then contain significantly more than the five renormalizable and four super-renormalizable terms we focused on in the present article.) 

\item Without ``projectability'' the Hamiltonian constraint would become a full super-Hamiltonian constraint.

\item The dynamical equation would then pick up additional terms depending on the gradient and the Hessian of the lapse. (The forcing term $F_{ij}$ would in general be significantly more complicated.) 

\item For both $\xi\neq0$ and $\xi=0$ the flat-space spin-2 tensor graviton propagator we have calculated would not be affected; while the spin-0 scalar graviton can either eliminated at the linearized level for $\xi\neq0$ under suitable fall-off conditions at spatial infinity, or collapses to a linear expansion/contraction mode for $\xi=0$ (in the specific gauge used in this article).

\item The Friedmann equations we have extracted would not be affected.  (The lapse $N$ is unity in any FLRW spacetime.)
\end{itemize}
Thus generalizing to arbitrary lapse might not be as difficult as initially envisaged~\cite{Horava}, and this is a worthwhile option to consider.

Turning to the future, the most obvious tests of our current variant of Ho\v{r}ava's model would come from the observational limits on Lorentz violations~\cite{Mattingly, Liberati, Planck}, but by inspection the model should also fall into the PPN framework, and specifically be subject to ``preferred frame'' effects~\cite{PPN} --- this will lead to stringent limits on the size of the Lorentz breaking parameter $\zeta$ arising from solar system physics.  Up to this stage we have not had to make any specific commitment as to how matter (in the form of elementary particles as opposed to cosmological fluid) couples to the gravitational field: Because the gravitational field is still completely geometrical, albeit with a ``preferred frame'', it seems to us that there is no intrinsic difficulty in maintaining a ``universal'' coupling to the gravitational field, and so there is no need for violations of the equivalence principle in this class of models --- we expect the universality of free fall to be maintained and to not see any likelihood for non-zero signals in E\"otv\"os-type experiments.

In conclusion we would argue that this is one of very few quantum gravity models that has any realistic hope of direct confrontation with experiment and observation, and that it is well worth a very careful look.

\section*{Acknowledgments}
TPS was supported by STFC and during early stages of this work by NSF grant PHYS-0601800. MV was supported by the Marsden Fund administered by the Royal Society of New Zealand. SW was supported by a Marie Curie Fellowship.  We wish to thank Anzhong Wang for pointing out some annoying typos.



\end{document}